\documentclass{article}
\usepackage{graphicx} 
\usepackage{authblk}
\usepackage{amsmath}

\title{A computational model for gender asset gap management with a focus on gender disparity in land acquisition and land tenure security}
\author{Oluwatosin Ogundare}
\author{Lewis Njualem}

\affil{California State University, San Bernardino}

\date{}

\begin{document}

\maketitle

\begin{abstract}
   Gender inequality is a significant concern in many cultures, as women face significant barriers to asset acquisition particularly land ownership and control. Land acquisition and land tenure security are complex issues that affect various cultural groups differently, leading to disparities in access and ownership especially when superimposed with other socio-economic issues like gender inequality. Measuring the severity of these issues across different cultural groups is challenging due to variations in cultural norms, expectations and effectiveness of the measurement framework to correctly assess the level of severity. While nominal measures of gender asset gap  provide valuable insights into land acquisition and tenure security issues, they do not fully capture the nuances of cultural differences and the impact of governmental and corporate policies that influence gender disparity in land ownership and control. The proposed framework aims to fill this gap by incorporating cultural and policy factors in developing a new measurement framework equipped with a more robust, comprehensive metric to standardize the approach to assessing the severity of gender asset disparity in a general sense but with a focus on land acquisition and tenure security to engender more effective interventions and policy recommendations.
\end{abstract}
\section{Introduction}
Women's land rights vary by region and country. Although progress has been made in some geographical areas to close the gender disparity in access to land acquisition and land tenure security, it remains a challenge in many parts of the world due to cultural and social norms. For example, in parts of Asia, male inheritance and control over land are prioritized, limiting women's ability to own and control land. In some parts of Sub-Saharan Africa, patriarchal customs and laws often restrict women's land rights, and they are dependent on male relatives for land access. Regardless, there have been efforts in many of these regions to promote women's land ownership, such as in Rwanda where women were granted equal inheritance rights in 1999 \cite{nyaya2016comparative}. There are many factors that contribute to observed gender disparity and without addressing some of the underlying concerns, the status quo is likely to continue. According to a study of a region in Tanzania, a country in sub-Saharan Africa, the main contributors to the gender inequality in land acquisition is illiteracy. Majority of the women are simply unaware of their entitlements or they lack the resources to challenge the status quo\cite{moyo2017women}. In the case of Zimbabwe, despite ratifying several conventions and declarations including establishing a legal framework governing women's land rights that prohibit discrimination against women in any sector, including agriculture and land ownership, the government has not taken sufficient steps to ensure that women are included in the development process and enjoy the right to legal tenure or non-discrimination when it comes to land ownership\cite{gaidzanwa2011women}. Essentially, the impact of policy in addressing the gender disparity problem is uneven i.e., similar policies might  have varying effectiveness in different geographical regions. As such, a framework for subjective policy impact quantification is needed. There are incentives associated with reduced gender asset disparity; Maetens et al. concluded that extending opportunity for land tenure security and land ownership for women improves agro-industrial processing in modern supply chains\cite{maertens2012gender}. Njualem et al. also supported this view that a decrease in gender asset disparity will improve the global sustainability score of supply chain networks in sub-Saharan Africa and regions of south Asia \cite{njualem1}\cite{9436751}. The focus of this paper is to introduce a mathematical formalism and an algorithm for the computation of a metric for evaluating the impact of government policies and agency action on minimizing the gender disparity gap.

\section{History of measures of gender disparity in land acquisition and land tenure security}
Gender disparity in land ownership is a phenomenon that has been studied by many scholars, especially when it concerns regions with cultural peculiarities such as sub-Saharan Africa and south Asia. It has become a resonating theme in contemporary global discourse, where efforts are being made to reduce inequalities and spur sustainable growth in areas of agriculture and supplemental primary products. These endeavors are in concert with critical Sustainable Development Goals (SDGs) as ratified by the United Nations, among which are \cite{rosen2017can}: 
\begin{description}
    \item [SDG \#2] End hunger, achieve food security, improve nutrition and promote sustainable agriculture
    \item [SDG \#5] Achieve gender equality and empower women and girls
    \item [SDG \#8] Promote sustained, inclusive, and sustainable economic growth, full and productive employment, and decent work for all
\end{description}  
There are several factors that contribute to gender based lack of access to equal rights and opportunities in asset acquisition and control. CD Deere et al. (2003)(2006) and Meinzen-Dick et al. (2014) discuss some of the factors in detail and propose a framework for quantifying the disparity \cite{deere2003gender} \cite{deere2006gender}\cite{meinzen2014gender}. The commonly identified gender related factors affecting asset distribution fall into four broad categories viz. socio-cultural, legislative, literacy and economic \cite{balas2022factors}.  Gender disparity is often quantified and presented in research using nominal values across gender categories. When the asset gap is wide, this statistic clearly illustrate the uneven distribution across the gender categories but fails to account for some of the thematic concerns. We find that the current approach is incapable of differentiating between two or more cases of gender disparity with similar or the same nominal statistics but have important differences. We are especially interested in the assessment of gender asset disparity immediately after the enactment of a government or corporate policy. A reduction in gender asset gap accrues over time and maybe impalpable at the onset of a policy. Therefore, there is a need to differentiate quantitatively between cases where active progress is being made and cases that are stagnant and no effort is being made to remedy the gender disparity problem when the nominal values do not yet reflect the impact, usually at the onset of the policy. This type of quantitative metric should be influenced by the thematic issues associated with the inequities of the gender asset disparity problem to be able to effectively reflect the impact of legislation and activism in this regard.

\section{Evaluating a quantitative metric}
First, we identify a set of criterion over which a suitable measurement framework should be evaluated and ultimately employ them in the development of a new metric to evaluate the gender disparity problem in land acquisition and land tenure security.

\begin{description}
   \item[Score Comprehension \& Interpretation]  A well defined score should span a scale that is easy to understand and whose values corresponding to a differentiable perceived state of the measured variable and can be used as a discriminator. For example, a score of 90\% on a test is an A, 95\% is an A+, 94\% is an A but it is perceived by an observer to be "almost" an A+.  
   \item[Score Composition] The score composition is a measure of linearly independent determinants accounted for in the score compared with the total number of linearly independent factors influencing the measured variable.
   \item[Score Resiliency] The score resiliency is the measure of a score's effective change due to a single point deviation or a single outlier. For example the Grade Point Average (GPA) is an effective measure of overall performance over a time period and is not susceptible to single point deviations when the time scale is large enough.
   \item[Score Consistency] The classical theory postulates that some degree of uncertainty and subjectivity is sometimes present in an individual's answer to a posed question. When a question, typically a question that involves a long-form response, is asked repeatedly to the same individual many times, a distribution of responses emerge. This distribution usually tend towards a central notion with some amount of variance \cite{Briggs2024-wi}, such that the following holds:\\ \\ 
   \begin{equation}
        Score_{recorded} = Score_{real} + \sum_{i} e_{i}
   \end{equation}
   \\ \hspace{15mm} Where $\sum_{i} e_{i}$ = Sum of combined errors from different sources.
\end{description} 
\section{The Sarafina Score}
We introduce the Sarafina score, named in honor of the South African fictional feminist icon, as a measure of gender asset  gap that incorporates policy effects as a stream of incentives. The Sarafina score rewards the enactment of new policies that address the gender disparity problem proportional to their effectiveness. Because the effect of policy intervention to address social problems trickle in, we expect that at the onset of a new policy the gender asset distribution may not be nominally different from the pre-policy state but we assert that a certain distinction has been achieved and should immediately reflect in a well defined assessment of the gender disparity problem. Hence, the policy component of the Sarafina score forecasts the expected effect of the policy on gender asset distribution and allows a temporal penalty to be assessed in the early stages or if the policy fails to yield the expected results over time. Intuitively, the penalty can be thought of as a function of the gap between the current state of the gender asset distribution and the projected state of the gender asset distribution. This allows us to assign value to policy activities at the beginning and adjust our expectation as time passes. Theoretically, the effective penalty iteratively approaches a limiting value but it is easier to think of it conceptually as going from maximum penalty to $0$ when the full effect of a policy is realized as shown in figure 1.

\begin{figure}
    \centering
    \includegraphics[width=1.0\linewidth]{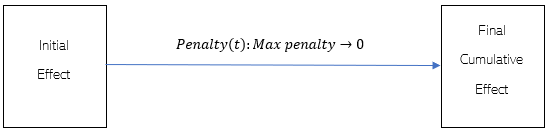}
    \caption{Conceptual diagram of penalty progression in time}
    \label{fig:1}
\end{figure}

Mathematically, the penalty is calculated as a function of the regret of the enacted government, corporate or cultural policy when compared to a implicit, maybe, unknown policy that would have achieved the desired reduction in gender disparity in asset distribution. The regret, $R(t)$ is such that at time, t:
\begin{multline}
    R(t) = \lvert \textit{ Observed Nominal Gap (\%)}
- \textit{Projected Nominal Gap (\%) } \rvert
\end{multline}
Consequently, the penalty function, $\phi(t)$, computes a cumulative relative regret as the ratio of the regret values over a period of time to the corresponding nominal gap of the gender disparity. $\phi(t)$ is approximated using a discrete formula as follows:
\begin{equation}
    \phi(t) \approx \frac{1}{k}\sum_{k=1}^{n} \frac{R[k]}{\textit{Nominal Gap}[k]} 
\end{equation}
\\Such that $\phi(t) \leq 1$ since the projected nominal gap is monotonically decreasing, $R[k]$ refers to the regret at discrete time step $k$. Therefore, computing the impact of policy contribution to minimizing the gender asset gap at a particular time, t is computed as follows:
\begin{equation}
    \textit{Policy Impact }= \hat{P}_{final} [1 - \phi(t)]
\end{equation}
\\$\hat{P}_{final}$ is the estimated or projected final reduced gender disparity (\%) due to the policy impact. The overall gender disparity score, a.k.a the Sarafina score at a particular time is simply the difference between the nominal gender asset disparity and the policy impact at that time. An effective policy should maintain its Sarafina score when evaluated at subsequent discrete time step, i.e., the average Sarafina score over any length of time should approach the same limiting value.
\begin{equation}
    \textit{Sarafina score}(t) = \textit{Nominal gender asset gap (\%)} - \textit{Policy Impact}\hspace{1mm} (\%)
\end{equation}
 \\When the projected Nominal Gap diverges from the observations then the Sarafina score begins to approach the observed Nominal Gap values. This signifies that the impact of the referenced governmental or organizational policy is not as effective as projected. The Sarafina score is designed to predict policy performance and weigh the effects of governmental and corporate action favorably against the gender gap problem even when the nominal values do not yet reflect strongly in that regard. Consequently, the Sarafina score is very sensitive to policy impact estimates and as such policy impact estimates cannot be arbitrarily computed. A sharply increasing Sarafina score might be an indication of poorly computed policy impact estimate or deliberate manipulation.
 
\section{Estimation of Policy Impact, $\hat{P}_{final}$ }
Predicting exact policy impact is not trivial \cite{ogundare2017you} and while we provide a simple computational strategy, it relies on the ability to collect and analyze historical data. Epstein et al. (1999) provided a regression model for predicting the impact of policy on minority voting \cite{epstein99}. Galleotti et al. (2019) used proxy indicators to track the effect of policy on pollution abatement \cite{GALEOTTI2020111052}. Ferrano et al. (2014) provided a stronger metric for estimating policy impact by computing the conditional statistical expectation as a measure of causal relationship between observed change and policy effect \cite{ferraro2014advances}. We propose an amalgamation of the ideas suggested by Galleoti and Ferrano by computing the statistical expectation of a policy outcome based on the distribution of a vector of proxy indicators. We create a simple but effective model by selecting, $C$ as a set of discrete improvement categories that can be used to classify or assigned to a vector proxy indicators. We choose $C = \{ C_{i} \}$ such that $C_{i} =$ \% reduction in gender disparity category. To illustrate, $C$ can be chosen as follows $C = \{2\%, 4\%, 6\%\}$. This allows us to simply calculate the prior conditional distribution of proxy indicators as follows:
\begin{equation}
    Pr(\textit{Proxy Indicators} | C_{i}) = \prod_{k=1}^{n} Pr(\textit{Proxy Indicator}_{k} |  C_{i})
\end{equation}
This of course requires that every proxy indicator,$\textit{Proxy Indicator}_{k}$, in the vector of proxy indicators is linearly independent. Ultimately, with a good reference historical data that establishes a relationship with the chosen proxy indicators and policy performance, we can estimate the posterior probability, $Pr(C_{i}|\textit{Proxy Indicators})$ over a discrete set of improvement categories ${C_{i}}$. On this basis, the projected policy impact on the disparity of gender asset distribution is chosen as the category (\% reduction in gap) that maximizes the posterior probability, i.e.,
\begin{equation}
    argmax \{ Pr(C_{i}|\textit{Proxy Indicators}) \}
\end{equation}
For many third world or developing countries, we have identified a list of proxy indicators that might be effective in predicting or estimating the impact of government legislation or corporate policy in reducing the gender asset gap.
\begin{itemize}
   \item Economic GDP
   \item Higher education Gender ratio
   \item Birth rate
   \item Domestic Violence Incidence - Investigation Ratio
   \item Judicial Effectiveness
\end{itemize}
\section{Predicting Policy Performance in Brazil and Mexico}
In Table 1, we present the distribution of land asset over gender in Brazil. Subsequently, in Table 2 we present similar data for Mexico for different years.

\begin{table}[htp]
    \centering
    \scalebox{0.82}{
    \begin{tabular}{ccccc}
         Year& Men (\% Owned) & Women (\% Owned) & Nominal Gender Asset Gap (\%) & Total\\
        $2000^{1}$ & 89 & 11 & 78 & n=39904\\
        $2006^{2}$& 89.8 & 10.2 & 79.6 & n=2779\\
        $2017^{2}$& 85.2 & 14.8 & 70.4 & n=2779\\
    \end{tabular}}
    \caption{Gender land asset distribution in Brazil}
    \label{tab:my_label}
\end{table}
$^{1}$ \cite{deere2003gender}, $^{2}$ \cite{araujo2024seeds}

\begin{table}[htp]
    \centering
    \scalebox{0.82}{
    \begin{tabular}{ccccc}
         Year& Men (\% Owned) & Women (\% Owned) & Nominal Gender Asset Gap (\%) & Total\\
        $1984^{3}$ & 87 & 13 & 74 & n=225\\
        $1996^{3}$& 78 & 22 & 56 & n=77\\
        $2002^{2}$& 77.6 & 22.4 & 55.2 & n=2.9m\\
    \end{tabular}}
    \caption{Gender land asset distribution in Mexico}
    \label{tab:my_label}
\end{table}
$^{3}$ \cite{hamilton2002neoliberalism}, $^{2}$ \cite{araujo2024seeds}
\\\\In this case study, we use the Sarafina score to predict the impact of Espaco Feminista, founded in 2008 on the gender disparity problem in land acquisition and land tenure security in Brazil. Similarly, we review the impact of the 1992 revision of the Mexican constitution to promote private influence in the agricultural industry \cite{hamilton2002neoliberalism} on the gender gap problem in land acquisition and land tenure security.
We set the estimated policy impact to be 25\% reduction in the nominal gender gap before the effect and we interpolate the data and present our very rough estimates for the implicit data for Brazil in Table 3. Mexico follows a similar pattern but converges more quickly as inferred from the data in Table 2.\\

\begin{table}[htp]
    \centering
    \scalebox{0.95}{
    \begin{tabular}{ccccc}
         Year& Nominal Gender Asset Gap (\%) & Sarafina Score (\%)\\
        $2008$ & 79.6 & 59.7 \\
        $2009$& 78.58 & 59.85 \\
        $2010$& 77.56 & 60.00\\
        $2011$ & 76.54 & 60.15 \\
        $2012$& 75.52 & 60.31 \\
        $2013$& 74.5 & 60.48\\
        $2014$ & 73.48 & 61.64 \\
        $2015$& 72.46 & 61.81 \\
        $2016$& 71.44 & 61.98\\
        $2017$& 70.42 & 61.15\\
    \end{tabular}}
    \caption{Gender land asset distribution in Brazil}
    \label{tab:my_label}
\end{table}

As shown in Figure 2, the Sarafina score uses the nominal gap data to compute a penalty against the estimated policy impact and adjusts the forecast of the ultimate impact of the policy.
\begin{figure}[hbt!]
    \centering
    \includegraphics[width=0.72\linewidth]{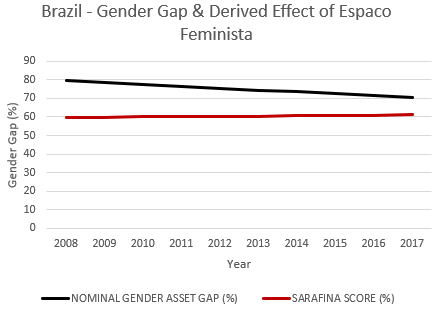}
    \caption{Convergence of the Sarafina score and the nominal gender asset gap distribution}
    \label{fig:enter-label}
\end{figure}

\section{Conclusion}
The Sarafina score is very effective in evaluating long term success of governmental, organization or cultural policies enacted to address the gender asset disparity problem and weigh the impact of these types of policies when assessing the severity of the gender gap problem.

\bibliographystyle{plain} 
\bibliography{refs} 
\end{document}